\documentclass{article}

\usepackage{arxiv}

\usepackage[utf8]{inputenc} 
\usepackage[T1]{fontenc}    
\usepackage{hyperref}       
\usepackage{url}            
\usepackage{booktabs}       
\usepackage{amsfonts}       
\usepackage{nicefrac}       
\usepackage{microtype}      
\usepackage{lipsum}
\usepackage{graphicx}
\usepackage{subcaption}

\title{Multi-Agent Actor-Critic\\
with Generative Cooperative Policy Network
}

\author{
  Heechang Ryu
    \\
  KAIST
   \And
 Hayong Shin \\
  KAIST
  \And
 Jinkyoo Park
   \thanks{All authors are with the Department of Industrial and Systems Engineering, Korea Advanced Institute of Science and Technology (KAIST), Daejeon, Republic of Korea (e-mail: rhc93@kaist.ac.kr; hyshin@kaist.ac.kr; jinkyoo.park@kaist.ac.kr).}\\
  KAIST
}

\begin{document}
\maketitle

\begin{abstract}
We propose an efficient multi-agent reinforcement learning approach to derive equilibrium strategies for multi-agents who are participating in a Markov game. Mainly, we are focused on obtaining decentralized policies for agents to maximize the performance of a collaborative task by all the agents, which is similar to solving a decentralized Markov decision process. We propose to use two different policy networks: (1) decentralized greedy policy network used to generate greedy action during training and execution period and (2) generative cooperative policy network (GCPN) used to generate action samples to make other agents improve their objectives during training period. We show that the samples generated by GCPN enable other agents to explore the policy space more effectively and favorably to reach a better policy in terms of achieving the collaborative tasks. 
\end{abstract}


\section{Introduction}
As modern engineering systems are composed of multiple sub-components interacting with each other in a complex way, it becomes challenging to understand the behavior of the system and operate it efficiently. Multi-agent system modeling is a useful tool for modeling the complex and collective behavior of the target system by modeling subsystems interacting with each other. The control problem of a multi-agent system has drawn much attention in the past years, due to the practical and potential applications in autonomous vehicles \cite{fax2004information,cao2013overview}, robotics \cite{corke2005networked,matignon2012coordinated}, logistics \cite{ying2005multi}, power grid \cite{dall2013distributed}, and resource management \cite{perolat2017multi}. Since the agents in multi-agent system lack full knowledge on dynamic environment and other agents’ strategies, learning decision-making strategy, \textit{policy}, in a multi-agent system is much more challenging than in a single agent system.

Deriving control policies for multi-agent system can be generally described as Markov Game (MG), an extension of Markov Decision Process (MDP) to a multi-agent system. In MG, given the global state, each agent takes an independent action to maximize its own payoff (\textit{i.e., accumulated reward}), while interacting with other agents. The interactions imply that the payoff of an individual agent and the evolution of global state are determined by the joint action of all the agents given a specific joint state. Depending on the relationships among the agents’ payoff, different equilibrium concepts are employed to agents’ equilibrium policy. However, it is generally difficult to derive analytically equilibrium strategies for a general MG because the optimality concept does not apply anymore in MG.

Many researchers have employed Reinforcement learning (RL) \cite{sutton1998reinforcement} approaches to derive policies for agents in various MG with different equilibrium concepts. This field of study is referred to as multi-agent reinforcement learning (MARL). For example, \cite{littman1994markov} proposed minimax Q-learning to solve the two-player zero-sum MG. \cite{hu1998multiagent} proposed Nash Q-learning to derive the Nash equilibrium strategies for two-player MG. \cite{littman2001friend} proposed friend-or-foe Q-learning to resolve convergence issues of Nash Q-learning. In addition, \cite{greenwald2003correlated} employed correlated equilibrium concept to derive the equilibrium strategies. These algorithms require each agent to compute equilibrium values of a stage-game defined at each state and use the computed equilibrium values to update the Q values of each agent. Solving the equilibrium value defined in each state means that this algorithm requires each agent to make decision while maintaining certain assumptions on other agents. This approach works only for very simple systems with discrete and low dimensional state and action space and usually does not guarantee the convergence since all the agents are learning concurrently using the data observed during game playing.

On the other hand, in a single agent system, deep reinforcement learning (DRL) algorithm has shown great successes in many complex tasks and even superior performances to human in video games \cite{mnih2015human}, robotics \cite{levine2016end}, and Go game \cite{silver2016mastering,silver2017mastering}. Motivated by such big successes in DRL, many researchers start to employ DRL approaches to controlling multi-agent systems, which is referred to as deep MARL. A deep neural network function approximator \cite{goodfellow2016deep} has shown good potential in effectively modeling complex interactions among multiple agents while incorporating dynamics and stochastic environment. However, due to the difficulty of imposing equilibrium concept among other agents, the majority of Deep MARL algorithms are designed to solve cooperative or team game, where the joint policy can be derived based on optimality principle, not equilibrium concept. This approach seeks a decentralized policy of each agent to achieve global objective when deriving a centralized policy is prohibited due to problem complexity, and this problem is equivalent to solving decentralized MDP (Dec-MDP) \cite{oliehoek2008optimal,kraemer2016multi,foerster2016learning,foerster2017counterfactual}. The core part of this problem is to design a consensus mechanism that allows each agent to understand the interactions among other agents by looking even partial state.

To make all the agents coordinate effectively in Dec-MDP with Deep MARL, different consensus mechanisms have been employed. The first approach is to train centralized Q-network capturing the relationships among agents’ actions and the global return. This approach uses the global Q-network to derive decentralized policies for agents. For example, \cite{foerster2017counterfactual} proposed a counterfactual actor-critic algorithm to derive the decentralized policy when the globally shared return is used. \cite{zhang2018fully} also proposed similar approach but they considered more complex situation where each agent need to infer the global return (i.e., the sum of return from all agents) from individually received return. To inference the global return based on local and individual return, \cite{zhang2018fully} employed parameter sharing among neighboring agents in the networked agents.

To model the interactions among agents, an approach of learning other agents’ policy has also been proposed. For example, \cite{lowe2017multi} proposed multi-agent version of deep deterministic policy gradient algorithm to derive agents’ policies for both cooperative and competitive MG. This approach allows each agent to train individual Q-network with its own individual reward while using other agents’ actions estimated from the policy networks for other agents. To infer other agents’ actions, each agent learns other agent’s policy during the training period. Although the authors insist that the algorithm can derive decentralized strategies for non-cooperative MG, this approach has limited performance because this algorithm does not guarantee to reach equilibrium strategy when solving non-cooperative game. That is, this algorithm just tries to find best response strategy given the other agents’ strategies trained from the data, without making sure that the trained policy networks are consistent among agents, the necessary condition for Nash equilibrium in non-cooperative game.

In this paper, we propose a multi-agent actor-critic algorithm with generative cooperative policy network (GCPN). This algorithm employing individual actor and critic networks can utilize individual reward for deriving a decentralized policy for each agent. This algorithm is mainly focused on improving the cooperation among agents in achieving a collaborative task. The cooperative policy can also be derived in the previously proposed algorithm \cite{lowe2017multi} by setting the identical (shared) reward for all the agents, but its performance is limited due to the failure of modeling contribution by each agent \cite{chang2004all}. To resolve this issue, we introduce GCPN that is trained to increase not its own return but other agent’s return. The generated samples from GCPN are then used to train individual actor and critic networks. We experimentally show that the samples generated by GCPN induce more consistent and efficient collaborative behaviors among agents.

\section{Background}
Generally, partially-observable Markov decision process (POMDP) have been widely used to model a single agent’s dynamic decision making under stochastic environment. In this paper, we consider partially-observable Markov game (POMG), an extension of POMDP to multi-agent system, to model the interaction of multiple agents with the environment. $\mathnormal{N}$-agents POMG is defined as follows; $\mathcal{S}$ describes a set of state for entire system. $\mathcal{O}_{i}$ is a set of observation for agent ${i}$, and each agent acquires a private and limited observation from the state through its own observation function ${o_i}:\mathcal{S}\mapsto\mathcal{O}_{i}$. $\mathcal{A}_{i}$ is a set of action for agent ${i}$. Each agent has a deterministic policy $\mu_{\theta_{i}}:\mathcal{O}_{i}\mapsto\mathcal{A}_{i}$ that maps the local observation to the action to take. We consider a deterministic policy that can be learned in a continuous and high-dimensional action space (Silver et al. 2014) for each agent, the policy that is suitable for controlling various physical systems. When all the agents take actions, each agent gets its own reward ${r}_{i}:\mathcal{S}\times\mathcal{A}_1\times\dots\times\mathcal{A}_N\mapsto\mathbb{R}$ from the environment and the global state $\mathcal{S}$ evolves to the next state according to the state transition model $\mathcal{T}:\mathcal{S}\times\mathcal{A}_1\times\dots\times\mathcal{A}_N\mapsto\mathcal{S}$. We assume that the initial state is determined by the initial state distribution $\rho:\mathcal{S}\mapsto[0,1]$. The agent $i$ aims to maximize its discounted return $R_i=\sum_{t=0}^{\infty} \gamma^t r_i^t$, where $\gamma$ is a discount factor.

\subsection{Deep Q-Networks (DQN)}
\newcommand{\argmax}{\arg\!\max}
Deep Q-Network (DQN), an RL approach to solve MDP \cite{mnih2015human}, aims to find an optimal policy $\pi$ that maximizes the expected return. DQN approximates the expected return using a deep neural network as $Q(s,a;\phi)\approx\mathbb{E}[R\arrowvert s^t=s,a^t=a]$. The parameter $\phi$ of $Q(s,a;\phi)$ is then optimized as one that minimizes the loss $\mathcal{L}$ defined as:
\begin{equation} \label{eq:1}
\mathcal{L(\phi)}= {\mathbb{E}_{s,a,r,{s'}\sim D}}[Q(s,a;\phi)-y]^2,
\end{equation}
where $y=r+\gamma{\max_{a'}}{Q'}({s'},{a'};\bar{\phi})$ is target computed by target network. In addition, $D$ is experience replay buffer, where $(s,a,r,{s'})$ is stored from the steps of episodes. Both experience replay buffer and target network are intended to stabilize learning. Once the optimum network parameter ${\phi^*}$ is computed, the optimal policy can be expressed as ${\pi^*}(s)={\argmax_a}Q(s,a;\phi^*)$.

\subsection{Deep Deterministic Policy Gradient (DDPG)}
A deep deterministic policy gradient (DDPG) \cite{lillicrap2015continuous}, as an actor-critic algorithm, aims to derive directly the deterministic policy network, referred to as an  \textit{actor}, $a_t=\mu_\theta(s_t)$ that maximize the expected return defined as:
\begin{equation} \label{eq:2}
\mathcal{J(\theta)}=\mathbb{E}_{s\sim\rho,a\sim\mu_\theta}[R]\approx\mathbb{E}_{s\sim\rho,a\sim\mu_\theta}[Q(s,a;\phi)]
\end{equation}
The Q-network in Equation \ref{eq:2}, referred to as a  \textit{critic}, is optimized like Equation \ref{eq:1} in DQN. The parameters $\theta$ of the policy network $\mu_\theta(s_t)$ are optimized using a gradient ascent algorithm with the computed gradient as follows:
\begin{equation} \label{eq:3}
\nabla_\theta\mathcal{J}(\theta)\approx\mathbb{E}_{s\sim D}[\nabla_\theta Q(s,a;\phi){\arrowvert_{a=\mu_\theta(s)}}]
         =\mathbb{E}_{s\sim D}[\nabla_\theta\mu_\theta(s){\nabla_a}Q(s,a;\phi)\arrowvert_{a=\mu_\theta(s)}]
\end{equation}
The gradient is computed using chain rule as the expected value of the product between the gradient of Q-network with respect to actions and the gradient of the policy network with respect to its parameters. It is called a deterministic policy gradient (DPG) \cite{silver2014deterministic}. By iteratively updating the parameters for the actor and critic networks, the algorithm induces the optimal policy network $a_t^*=\mu_{\theta^*}(s_t)$.

\subsection{Multi-Agent Deep Deterministic Policy Gradient (MADDPG)}
Recently, MADDPG \cite{lowe2017multi}, an extension of DDPG to multi-agent system, has been proposed to solve POMG. Instead of optimizing a single actor network as in DDPG, MADDPG provides a framework that each agent derives an individual policy network, actor, $a_i=\mu_i (o_i;\theta_i)$ mapping the local observation $o_i$ to the action $a_i$ in order to maximize the expected return that is approximated as an individual Q-network $Q_i$, critic. MADDPG allows each agent to receive its own individual reward signal $r_i(s,a_1\dots,a_N)$. In this setting, the core part of MADDPG are (1) to learn each player’s critic network with local reward $r_i$ and (2) to derive the decentralized actor network using the individual critic network.

The parameter $\phi_i$ of Q-network $Q_i$ for each agent $i$ is optimized as minimizing the loss $\mathcal{L}_i$:
\begin{equation} \label{eq:4}
\mathcal{L}_i(\phi_i)={\mathbb{E}_{\mathbf{o},a,r,{\mathbf{o}'}\sim D}}[(Q_i(\mathbf{o},a_1\dots,a_N;\phi_i)-y_i)^2],
\end{equation}
where $\mathbf{o}=(o_1,\dots,o_N)$ is the observations of all agents and $y_i=r_i(\mathbf{o},a_1,\dots,a_N)+\gamma {Q'_i}({\mathbf{o}'},{a_1'},\dots,{a_N'};{\bar{\phi_i}})\arrowvert_{{a'_j}={\mu_i^j}({o'_j})}$ 
is target computed by critic target network with slowly updating parameter $\bar{\phi_i}$. In addition, the experience replay buffer $D$ contains $(\mathbf{o},a_1,\dots,a_N,r_1,\dots,r_N,{\mathbf{o}'})$.  Note that, in the next observation ${\mathbf{o}'}$, agent $i$ estimates which action will be taken by other agents by using inferring policy networks for other agents ${a'_j}={\mu_i^j} ({o'_j};{\theta_i^j})$ that is being trained by agent $i$ for $j=1,\dots,N$. In short, the individual critic network captures the influences of other agents’ joint action, affecting the future accumulated reward of agent $i$.

 The parameter $\theta_i$ of policy network for agent $i$ is optimized using a gradient ascent algorithm with the computed gradient as follows:
\begin{equation} \label{eq:5}
\nabla_{\theta_i}\mathcal{J}(\theta_i)
=\mathbb{E}_{\mathbf{o},a\sim D}[\nabla_{\theta_i}\mu_i(o_i)\nabla_{a_i}Q_i(\mathbf{o},a_1,\dots,a_N;\phi_i )\arrowvert_{a_i=\mu_i(o_i;\theta_i)}] 
\end{equation} 
Furthermore, for stabilizing learning, each agent has multiple sub-policy networks for policy ensembles. By iteratively updating the parameters for the actor and critic networks, the algorithm induces the optimal policy network $a_i^*=\mu_i^*(o_i)$.

\subsection{Consensus Mechanism}
Decentralized control aims to express a joint policy for all agents as a product of individual policies of the agents. When decentralized policies are properly derived, each agent’s local action determined solely based on the local observation can lead the global equilibrium among agents, i.e., coordination for the cooperation game. Consensus mechanism refers to the procedure of deriving the equilibrium among all the agents’ policies through their individual goal-seeking behaviors. It means that each agent’s greedy behavior induces the global cooperation among all the agents for a cooperative game.

There are several ways of embedding consensus among agents into local policies of agents during training. Such procedures are referred to as “ \textit{centralized training and decentralized execution}”. One of widely used approaches is to learn other players’ policies and use these policies to estimate the other players’ actions, the case of MADDPG. As a result, the individual Q-network and the local policy derived from the Q-network implicitly embed other players’ behaviors. When such individual policies are used, then the decentralized agents are subject to coordinate with each other to achieve the global joint objective.

Another approach is to train a global Q-network to capture influences of the agents’ joint action on the global return \cite{foerster2017counterfactual,zhang2018fully} and uses this Q-network to update the local policy. The counterfactual actor-critic algorithm \cite{foerster2017counterfactual} trains a single global Q-network, while FDMARL \cite{zhang2018fully} trains the individual Q-network with individual reward, and makes the Q-networks be equal by forcing them to share the parameters of the Q-networks among neighboring agents as:
\begin{equation} \label{eq:6}
\phi^i\leftarrow\sum_{j\in\mathcal{N}} c(i,j)\cdot\phi^i 
\end{equation} 
where $\mathcal{N}$ is a set of all agents, and $c(i,j)$ is an element of doubly stochastic weight matrix for transmitting the information from $i$ to $j$ in a communication network. 

\section{Methods}
We propose a multi-agent actor-critic algorithm with GCPN to derive collaborative strategies for POMG participants. We are specifically focused on deriving decentralized policies for agents to maximize the performance of the collaborative task conducted by all the agents.

\begin{figure}[ht]
  \centering
  \includegraphics[scale=0.4]{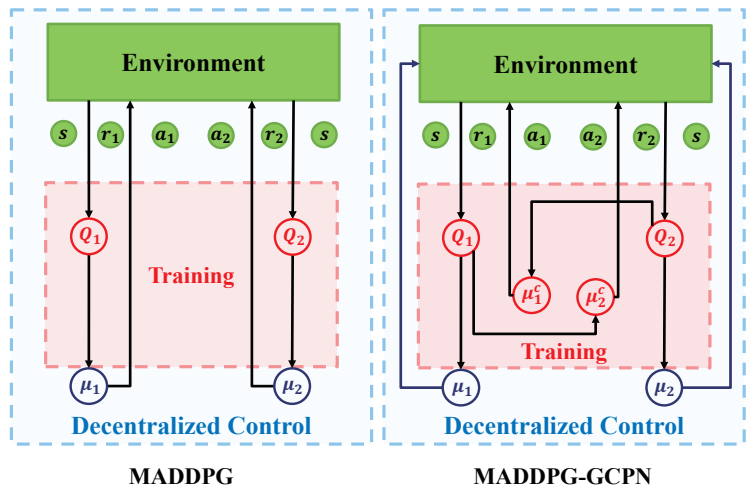}
  \caption{MADDPG (left) and MADDPG-GCPN (right).}
  \label{fig:fig1}
\end{figure}

The proposed algorithm is the variant of MADDPG in that it maintains the decentralized policy network and the individual Q-network. However, it is different from MADDPG in that it uses an additional policy network called  \textit{GCPN}. We refer to this algorithm as  \textit{MADDPG-GCPN}. In MADDPG-GCPN, each agent $i$ has the GCPN $\mu_i^c$ to maximize the return of the other agent apart from the original decentralized policy $\mu_i$ that is designed to maximize its own return. Thus, each agent has two different actor networks: (1) the behavioral actor network, GCPN, for generating samples and (2) greedy actor network for choosing actions for maximizing its own return. The difference between MADDPG and MADDPG-GCPN is shown in Figure \ref{fig:fig1}. To embed the cooperative behavior among agents into decentralized policy for each agent, we employ a procedure of centralized training and decentralized execution. The training of MADDPG-GCPN consists of training the followings:
\begin{itemize}
\item  \textit{Individual Q-network (critic)}: each agent trains individual Q-network, which is used to update the individual greedy policy for choosing optimal action during execution period, and the individual behavioral policy, GCPN, for generating action samples for exploration during training period.
\item  \textit{Individual greedy policy network}: similar to MADDPG, individual policy is trained to maximize the expected return represented as a individual critic network.
\item  \textit{Generative cooperative policy network}: the GCPN $\mu_i^c$ generates action samples to interact the environment during training period for learning the actor and critic networks..
\end{itemize}
	   
In execution period after training, the Q-networks and GCPNs of all agents are removed. Only the individual greedy policy of each agent controls the action based on local observation.

In MADDPG-GCPN, the parameter $\phi_i$ of the individual Q-network of each agent $i$ is optimized as minimizing the loss $\mathcal{L}_i$:
\begin{equation} \label{eq:7}
\mathcal{L}_i(\phi_i)={\mathbb{E}_{\mathbf{o},{a^c},r,{\mathbf{o}'}\sim {D^c}}}[(Q_i(\mathbf{o},a_1^c,\dots,a_N^c;\phi_i)-y_i)^2,
\end{equation}
where $y_i=r_i(\mathbf{o},a_1^c,\dots,a_N^c)+\gamma {Q'_i}({\mathbf{o}'},{a_1^{c'}},\dots,{a_N^{c'}};{\bar{\phi_i}})\arrowvert_{{a_j^{c'}}={\mu_j^{c'}({o'_j})}}$ 
is the target computed by target network. The experience replay buffer $D^c$ contains $(\mathbf{o},{a_1^c},\dots,{a_N^c},r_1,\dots,r_N,\mathbf{o'})$ where the $a_j^c$ is generated not from greedy actor but from the GCPN. In the target network in Equation \ref{eq:7}, the actions of other agents, except agent $i$, are estimated by using additional inferring policy networks ${a_j^{c'}}={\mu_j^{c'}}({{o'}_j})$.

Then, the parameter $\theta_i$ of individual greedy policy $\mu_i$ of each agent $i$ is optimized using a gradient ascent algorithm with the computed gradient as follows:
\begin{equation} \label{eq:8}
\nabla_{\theta_i}\mathcal{J}(\theta_i)
=\mathbb{E}_{\mathbf{o},{a^c}\sim {D^c}}[\nabla_{\theta_i}\mu_i(o_i)\nabla_{a_i}Q_i(\mathbf{o},{a_1^c},...,{a_i},...,{a_N^c};\phi_i )\arrowvert_{a_i=\mu_i(o_i;\theta_i)}] 
\end{equation} 
The individual greedy policy is trained in a similar way to MADDPG.

The major contributions of this study are introducing additional policy network for generating samples and suggesting the training method for that network.  
The parameter $\theta_i^c$ of the GCPN $\mu_i^c$ of each agent $i$ is optimized using a gradient ascent algorithm with the specially computed gradient as follows:
\begin{equation} \label{eq:9}
\nabla_{\theta_i^c}\mathcal{J}(\theta_i^c)
=\mathbb{E}_{\mathbf{o},{a^c}\sim {D^c}}[\nabla_{\theta_i^c}\mu_i^c(o_i)\nabla_{a_i^c}Q_{-i}(\mathbf{o},{a_1^c},...,{a_N^c};\phi_i )\arrowvert_{a_i^c=\mu_i^c(o_i;\theta_i^c)}] 
\end{equation} 
where ${-i}$ is the agents in $N$-agents excluding agent $i$. The policy gradient for the GCPN $\mu_i^c$ is computed using (1) the gradient of Q-network of other agents with respect to actions and (2) the gradient of GCPN with respect to its parameters. The trained GCPN then tends to generate action samples that can induce a higher returns of other agents, which is the reason why we call this as cooperative policy network.

For stabilizing training, the agent $i$ has $\arrowvert-i\arrowvert$-GCPNs, where $\arrowvert-i\arrowvert$ is the cardinality of ${-i}$, and each GCPN is trained from one agent of ${-i}$. Since each agent has several GCPNs, it generates a sample by randomly selecting one of the GCPNs at every step or considering the GCPN ensembles. In the step of updating the parameters, both the GCPN and the individual greedy policy are learned while only the GCPNs are considered in generating samples. Furthermore, there are $K$ different sub-policies for the GCPN and individual greedy policy. There are experience replay buffers $D_i^{c,k}$ for each sub policy of the GCPNs.

In addition, the consensus mechanism through GCPNs preserves the privacy among the agents because it uses own local reward and the gradient from the critics of the other agents during learning and uses only own decentralized policy and local observations when executing.

\section{Results}
\begin{figure}[ht]
  \centering
  \fbox{\includegraphics[scale=0.4]{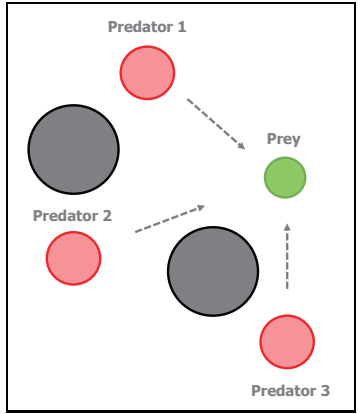}}
  \caption{Predator-prey in OpenAI Gym.}
  \label{fig:fig2}
\end{figure}

To evaluate the performance of the proposed algorithm, the MARL task in OpenAI Gym \cite{brockman2016openai} is used as a target POMG.  \textit{Predator-prey} is a task where several predators can gain higher reward,  \textit{score}, when they cooperate with each other to catch a prey. Figure \ref{fig:fig2} illustrates the game environment with three predators and one prey with two landmark obstacles. 

In the game, the predators should cooperate with each other to catch the prey because the prey is faster than predators. The predator-prey game is defined as the following POMG:
\begin{itemize}
\item The partially-observable state containing the relative positions, velocities of the agents, and the positions of the landmarks. 
\item The action of each agent is a movement toward all directions.
\end{itemize}
	
We consider the two reward function structures:
\begin{itemize}
\item Shared reward function ${r_i=10}$ if  \textit{any} agent catch the prey, otherwise $0$ for $i=1,2,3$.
\item Individual reward function ${r_i=10}$ if agent i catch the prey, otherwise $0$ for $i=1,2,3$.
\end{itemize}
	
That is, we consider two POMGs, one with shared reward structure and the other with individual reward structure. For these two different POMGs, we will compare the performances of different MARL approaches in terms of maximizing the winning rate of the predators.

We compare the performances of the following four learning algorithms:
\begin{itemize}
\item CF \cite{foerster2017counterfactual}: the MARL actor-critic algorithm designed for a team game with a shared return for all agents. Each agent who is jointly maximizing its shared return through single global Q-network directly induces the collaborative behavior among the agents. Because of this model assumption, CF applies to only the shared reward case. In this experiment, CF indicates that only single global Q-network with MADDPG is used.
\item FDMARL \cite{zhang2018fully}: the MARL actor-critic algorithm designed for a cooperative game with individual return for each agent. The algorithm requires each agent to infer the global return through parameter sharing and uses the inferred global return to derive the decentralized policy. In this experiment, FDMARL indicates that only parameter sharing with MADDPG is used.
\item MADDPG \cite{lowe2017multi}: the MARL actor-critic algorithm designed for general non-cooperative game. This algorithm requires each agent to learn other players’ policies and use the trained policies to approximate its own Q-network. In addition, the individual policy is also trained to interact with the individually updated Q-network.
\item MADDPG-GCPN1: the proposed algorithm with the randomly selected GCPNs in sample-generating.
\item MADDPG-GCPN2: the proposed algorithm with GCPN ensembles in sample-generating.
\end{itemize}
	
In performing tasks, the prey is learned by DDPG because it is a single agent. 

\begin{figure}[t]
  \centering
  \begin{subfigure}[t]{0.475\textwidth}
        \includegraphics[scale=0.22]{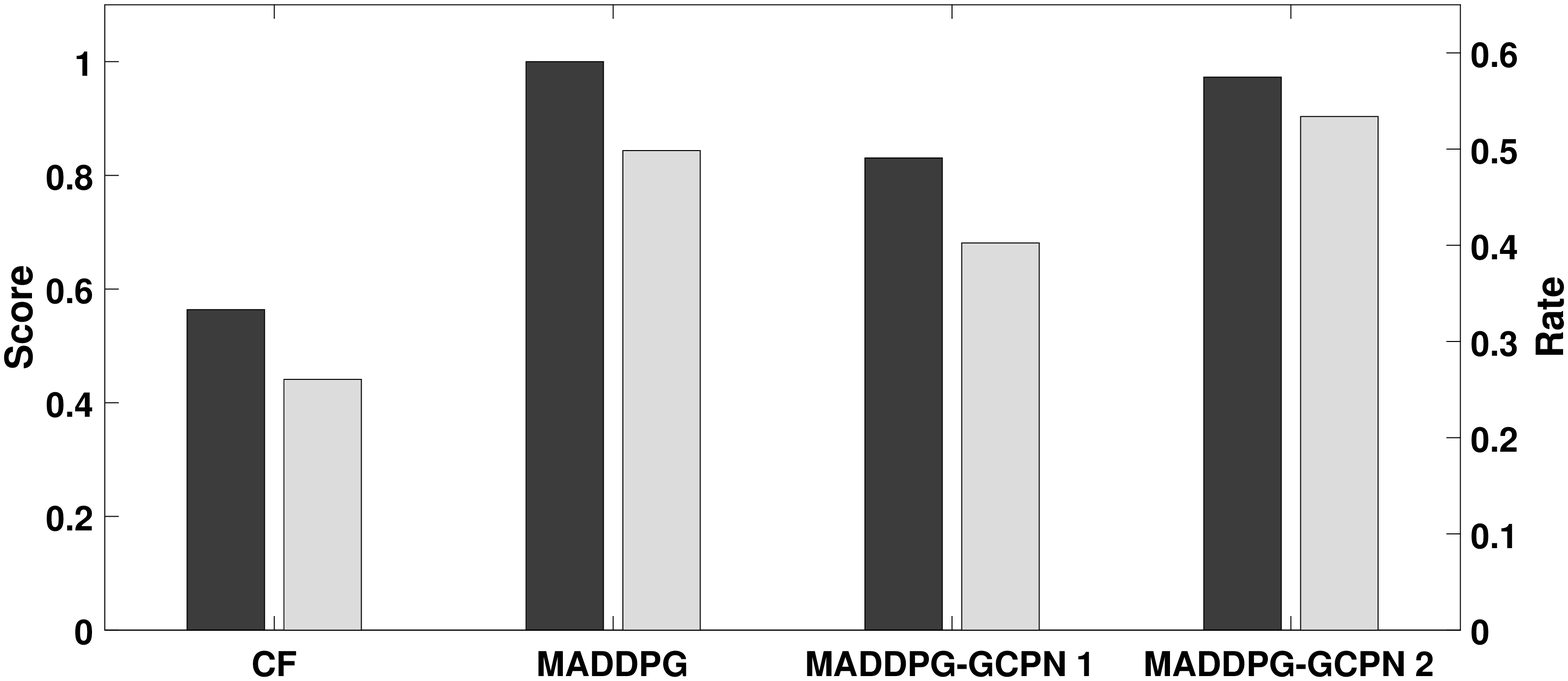}
        \caption{}
        \label{fig:label3:(a)}
  \end{subfigure}
  \begin{subfigure}[t]{0.475\textwidth}
        \includegraphics[scale=0.22]{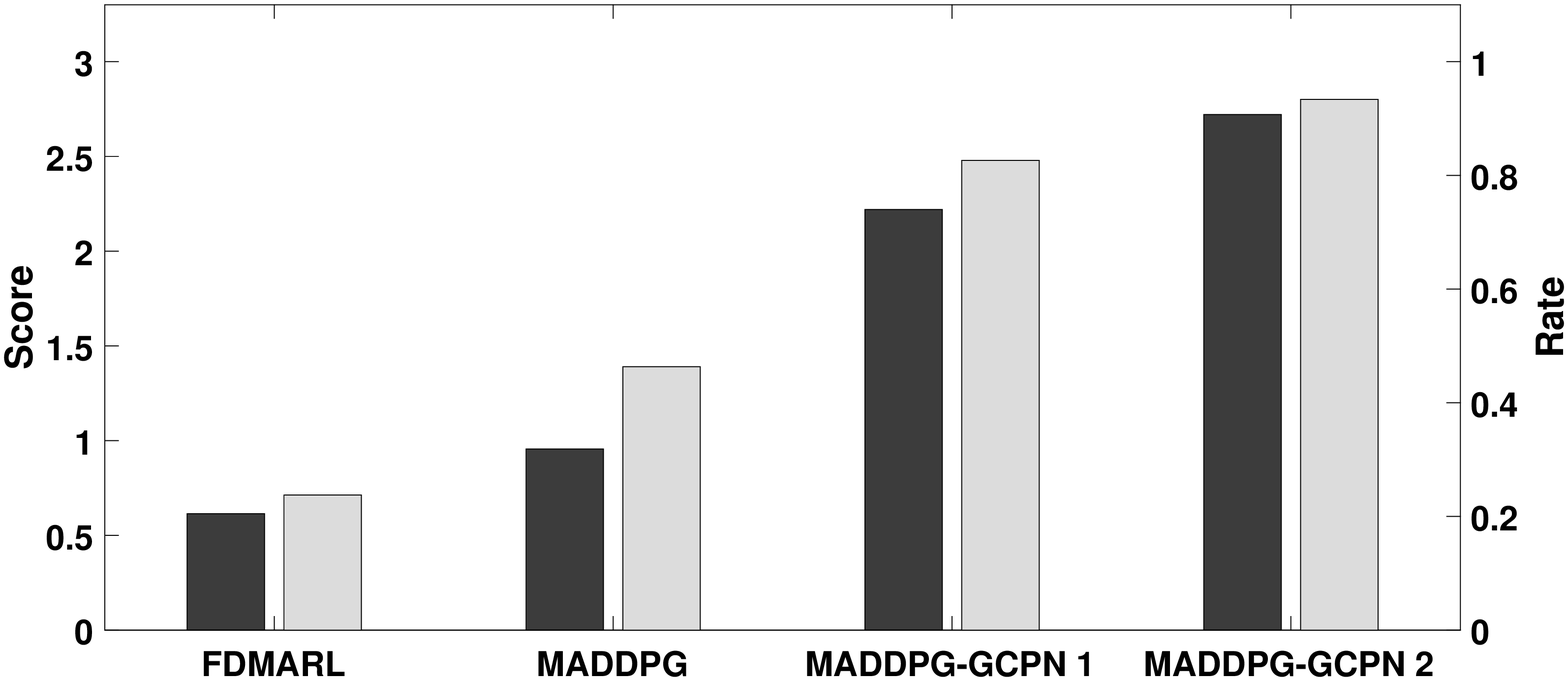}
        \caption{}
        \label{fig:label3:(b)}
  \end{subfigure}
  \caption{Normalized score (dark) and catching rate (light) in the shared (a) (left) and individual (b) (right) reward cases with each algorithm.}
  \label{fig:fig3}
\end{figure}

In Figure \ref{fig:label3:(a)} for the shared reward case and Figure \ref{fig:label3:(b)} for the individual reward case, they show the normalized score and catching rate of the predators with each algorithm. The scores of predators are normalized based on the MADDPG score. The catching rate is the rate at which the prey is caught by the predators. In the shared reward case in Figure \ref{fig:label3:(a)}, each algorithm has similar scores although the score of CF is slightly lower. Although the algorithms have the consensus mechanism for cooperation, the shared reward leads to artificial and weak cooperation in training period, preventing the agents from reaching equilibrium.

\begin{figure}[t]
  \centering
  \begin{subfigure}[t]{0.475\textwidth}
        \includegraphics[scale=0.22]{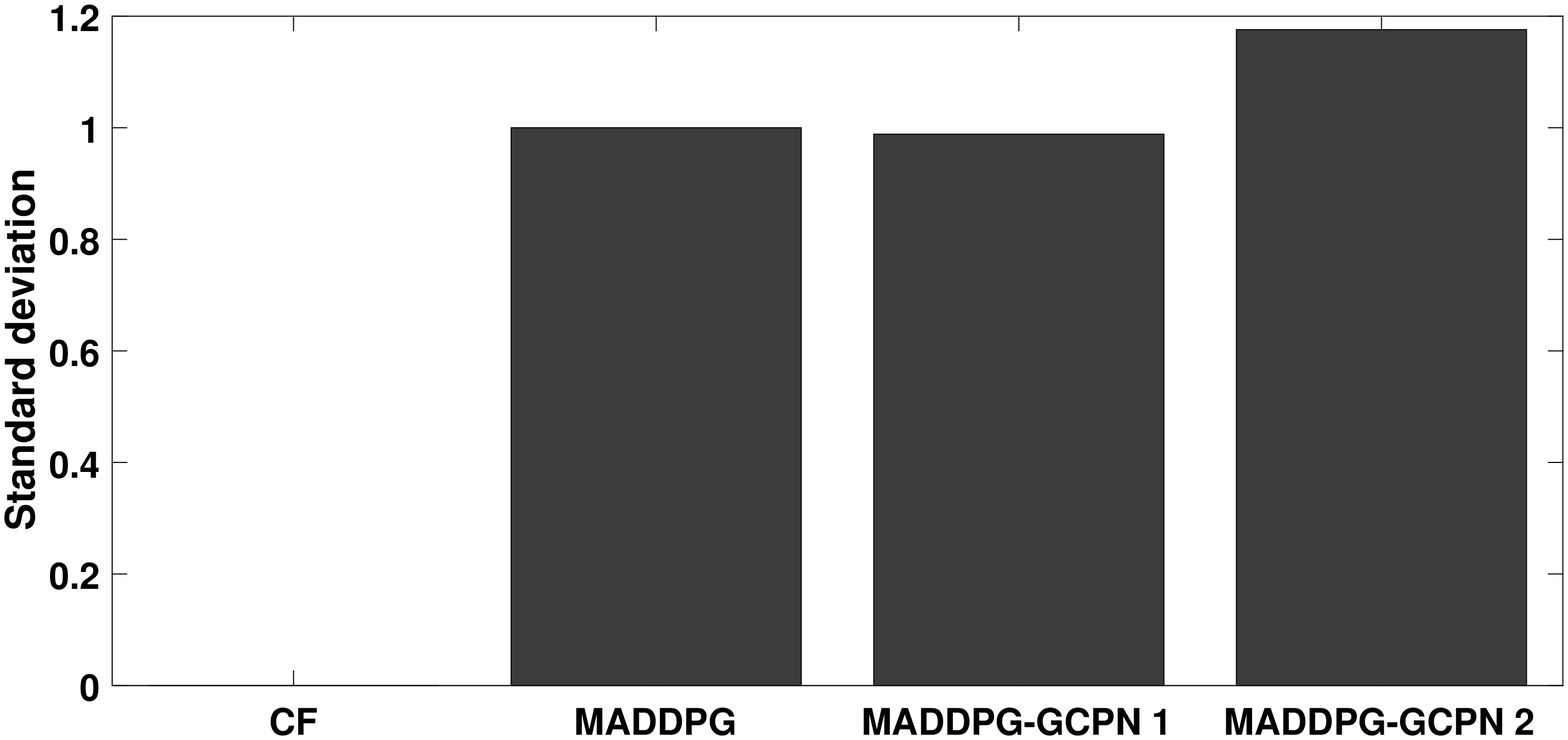}
        \caption{}
        \label{fig:label4:(a)}
  \end{subfigure}
  \begin{subfigure}[t]{0.475\textwidth}
        \includegraphics[scale=0.22]{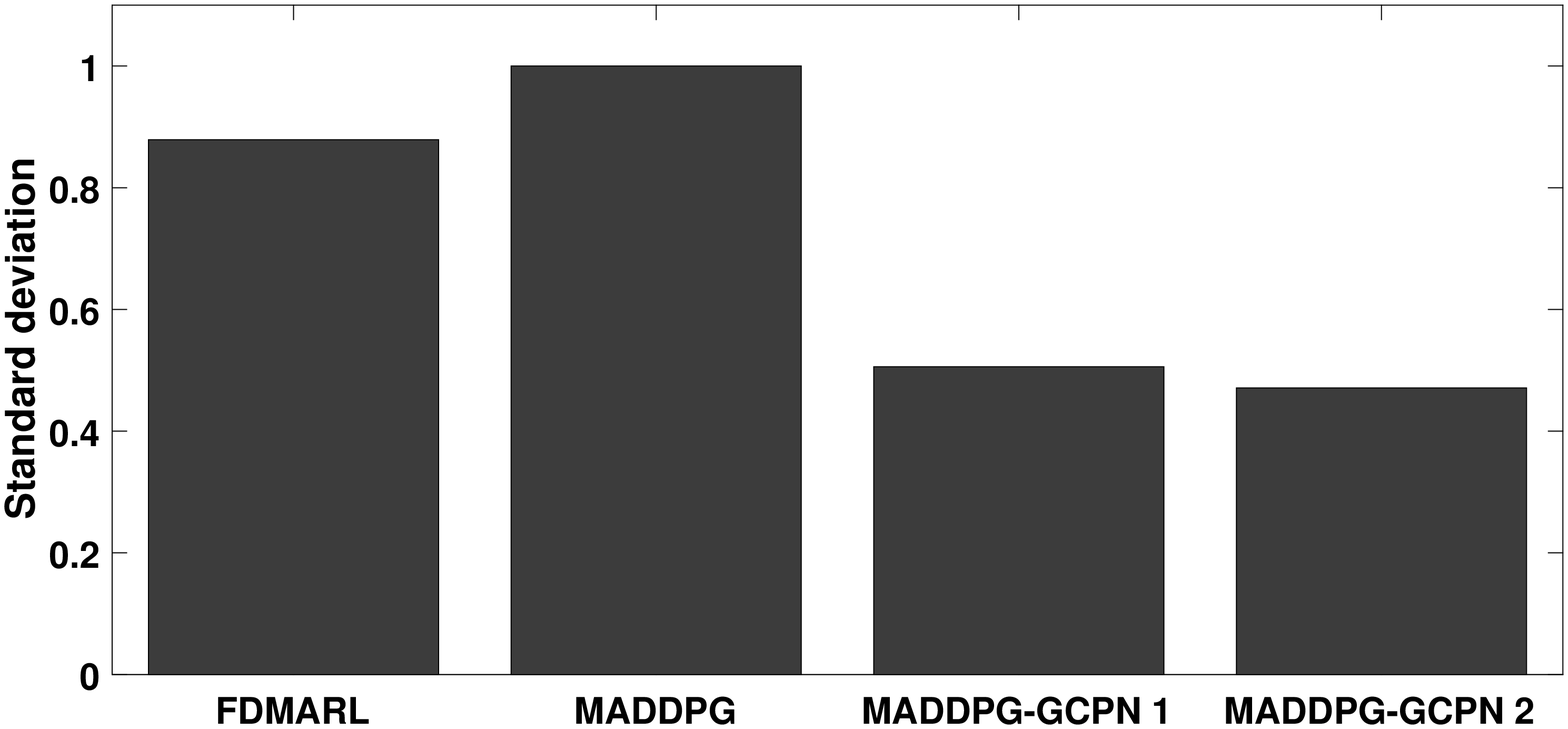}
        \caption{}
        \label{fig:label4:(b)}
  \end{subfigure}
  \caption{Normalized average standard deviation of trained critics in the shared (a) (left) and individual (b) (right) reward cases with each algorithm.}
  \label{fig:fig4}
\end{figure}

In the individual reward case in Figure \ref{fig:label3:(b)}, the normalized scores tend to be very different. In particular, MADDPG-GCPNs have higher scores than other algorithms. This is because when the agents receive individual rewards, they try to perform various actions for each return of others through GCPN, leading to more optimal training and equilibrium. On the other hand, FDMARL and MADDPG also lead cooperation in the case of individual return because of the existence of the consensus mechanism for cooperation. However, due to the anthropogenic consensus mechanism, the agents in the algorithms are not learned to achieve equilibrium for cooperation. 

Being trained optimally for cooperation is seen in the individual Q-network values of agents. In most cases, homogeneous agents exist in tasks requiring the cooperation of agents. The homogeneous agents are considered to have homogeneous actors and critics if they are trained optimally. In this consideration, for the algorithms evaluated in Figure \ref{fig:label3:(a)} and \ref{fig:label3:(b)}, the critic values of the cooperating agents are calculated, and their standard deviation is shown in Figure \ref{fig:fig4}. In the figure, the CF is excluded because it has global Q-network, not individual. The values of Figure \ref{fig:fig4} are normalized based on the MADDPG. As Figure \ref{fig:label4:(a)} shows, the average standard deviation of the algorithms is similar in the shared reward case. However, in Figure \ref{fig:label4:(b)}, the MADDPG-GCPNs show low standard deviations in the individual reward case while the other algorithms show high standard deviations. This may suggest that the agents working in MADDPG-GCPNs are trained to behave more homogeneously.

\section{ESS Control Application}
The MADDPG-GCPN applies to the real-world problem of controlling energy storage systems (ESSs) in several microgrids. The microgrids are decentralized and localized energy distribution systems consisting of energy generators, communities of households, and ESSs. Although renewable energy generators, such as wind turbines, which can function as independent energy sources, reduce the dependence on the central grid, such as energy company, it is difficult to operate reliably due to its intermittent energy generation of renewable energy subject to stochastic variations of environmental conditions. 

\begin{figure}[ht]
  \centering
  \includegraphics[scale=0.4]{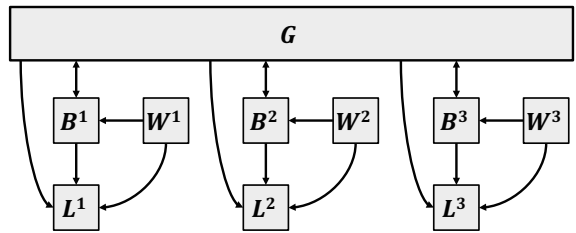}
  \caption{Energy company and three microgrids.}
  \label{fig:fig5}
\end{figure}

An ESS can charge energy when renewable energy is over-produced or when the energy price of the energy company is low, and it can discharge the energy when the demand for energy from the community is high or when the energy price of the energy company is high. When optimally planned, an ESS can relieve the imbalance between energy production and consumption on the microgrid, leading to reliable operation of the microgrid. Generally, the energy price depends on the energy load on the energy company, and the higher the load, the higher the price.  The load on the energy company arises from the energy transactions of microgrids and the energy company. In Figure \ref{fig:fig5}, the nodes show energy company $G$ and microgrids with wind turbines $W^i$, communities $L^i$,  and ESSs $B^i$, and there are three microgrids. 

The problem of controlling multiple ESSs is the extension of single ESS control problem \cite{atzeni2013demand}. Because the energy price is influenced not only by its own microgrid but also by other microgrid, each microgrid must control the ESS by considering other microgrids. Therefore, this problem can be modeled as a multi-agent system and solved by our algorithm. We consider the ESS of each microgrid as an agent and aim to control the energy charge and discharge schedule between the ESS and each node with the optimal policy. The energy demands from the communities and wind energy generated from the wind turbines is provided by actual data from several European countries. Controlling distributed ESSs in a decentralized manner can be formulated as the following MG:
\begin{itemize}
\item State contains the past trajectory of energy demand, generated wind energy, energy price, and residual energy in ESS.
\item Action is energy flows $\mathcal{X}_t$ between the ESSs and each node in Figure \ref{fig:fig5}.
\item Energy cost is defined as $C_t C_t^i$ if the total load of all microgrids is $C_t$ and the load of each microgrid $i$ is $C_t^i$ at time t. If the sum of energy consumptions of all agents in a specific time is large, the energy price increases, and thus all the players should pay more energy cost.
\end{itemize}

The goal of each agent is to minimize the accumulated energy cost over an infinite horizon.

We employ MADDPG-GCPN to solve the MG. The individual Q-network approximates the negative energy cost of each agent and, as suggested, two policy networks are trained for exploration and exploitation. After training, each ESS agent observes the past historical state and choose the ESS control actions by using individual decentralized policy network.

\begin{figure}[t]
  \centering
  \begin{subfigure}[t]{0.475\textwidth}
        \includegraphics[scale=0.22]{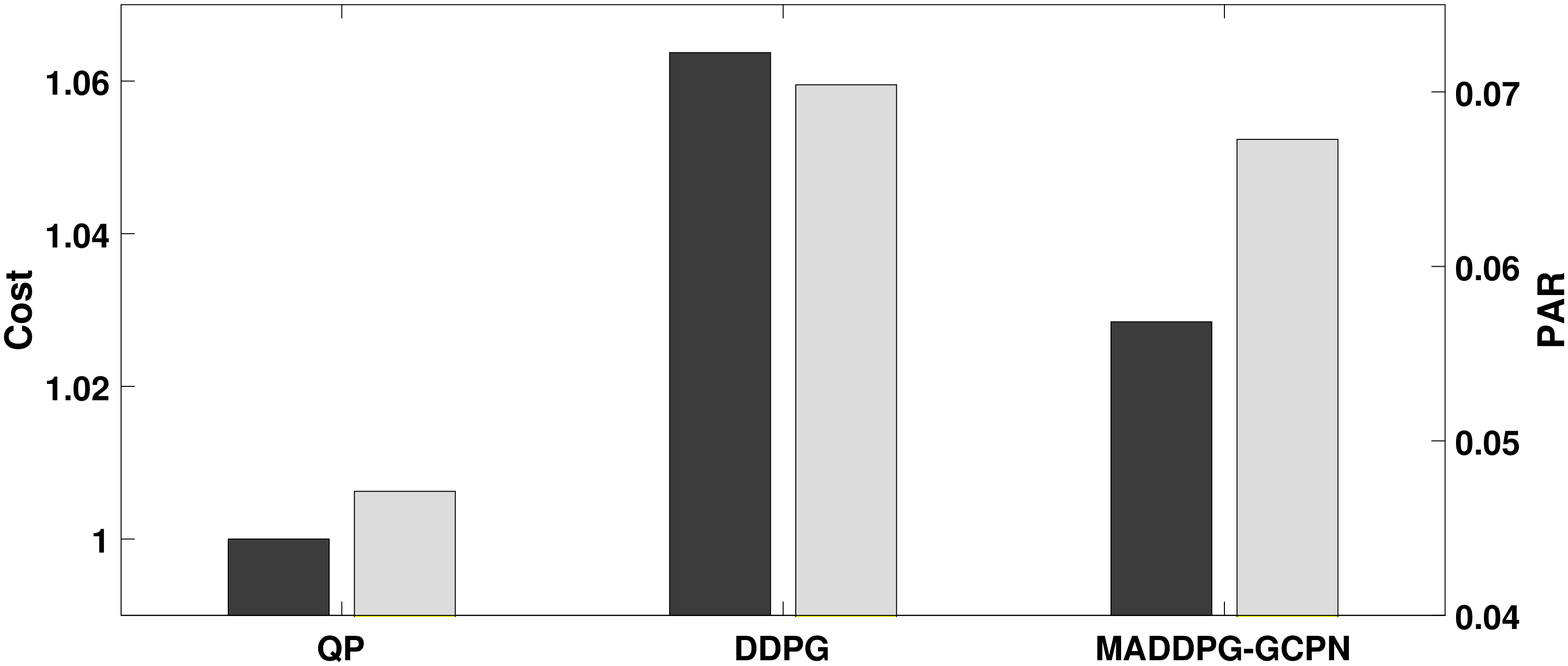}
        \caption{Normalized cost (dark) and PAR (light) with each algorithm.}
        \label{fig:label6:(a)}
  \end{subfigure}
  \begin{subfigure}[t]{0.475\textwidth}
        \includegraphics[scale=0.235]{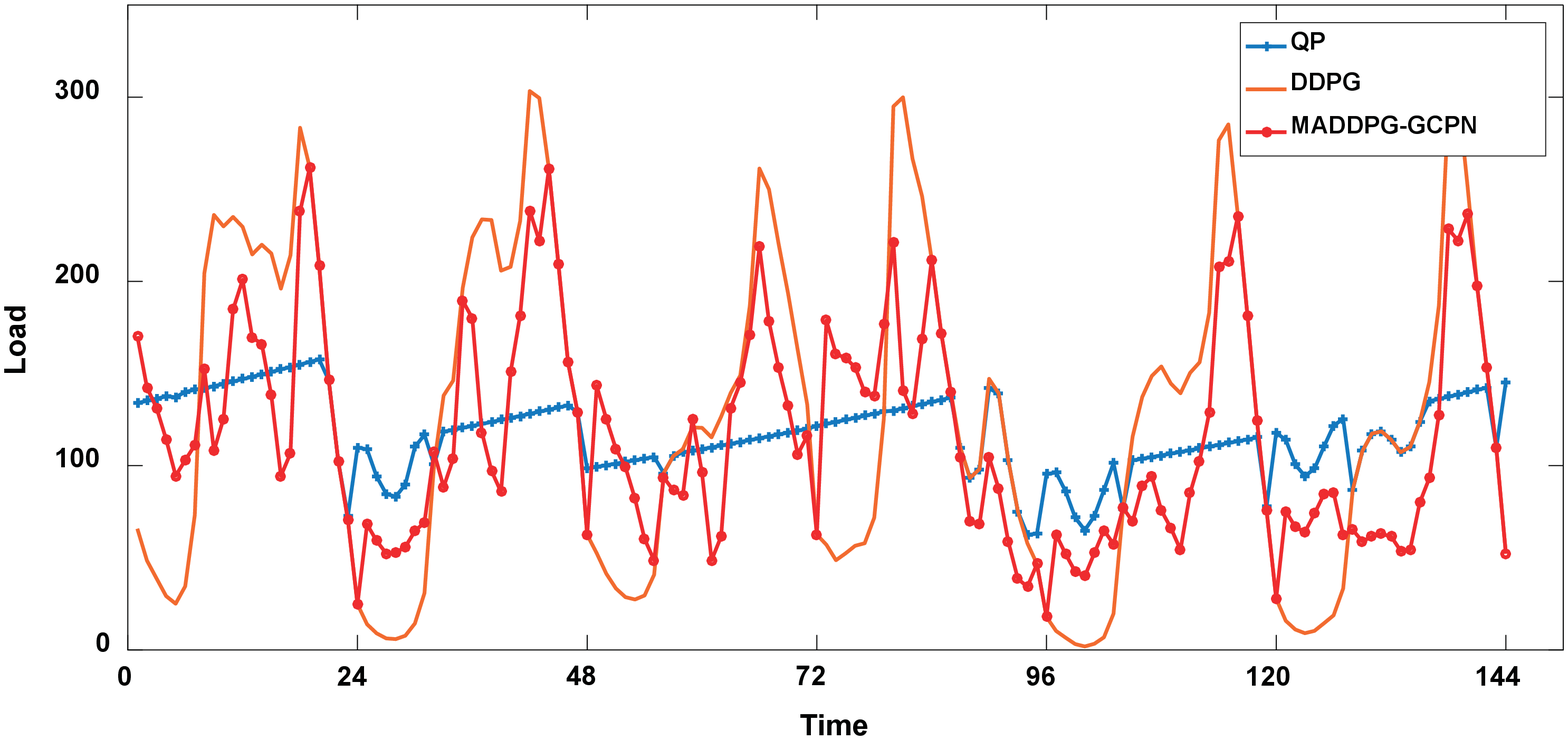}
        \caption{Total energy consumption of all microgrids.}
        \label{fig:label6:(b)}
  \end{subfigure}
  \caption{}
  \label{fig:fig6}
\end{figure}

Figure \ref{fig:label6:(a)} compares the results when three methods are employed:
\begin{itemize}
\item Quadratic programming (QP) approach: the joint actions for the three ESSs are optimized assuming there is a centralized controller that knows full knowledge about the future demand and energy generation and the system’s dynamics. 
\item DDPG approach: Ignoring the interactions among ESSs, each DDPG algorithm is focused on training individual actor and critic networks.
\item MADDPG-GCPN: the actor networks trained by the proposed method are employed.
\end{itemize}
	
The cost in Figure \ref{fig:label6:(a)} is the energy cost of all microgrid accumulated for 365 days of test data. The total energy costs are then normalized based on the cost computed by QP for performance comparison. The PAR is a peak-to-average ratio, which is a relative peak to average load.

DDPG results in high cost because each agent controls its ESS considering only its own  reward without considering the interactions among ESSs’ operation. On the other hand, MADDPG-GCPN results in a cost closer to the cost of QP. This is because each agent tries to reduce the hourly total energy consumption by all agents through coordination, as shown in Figure \ref{fig:label6:(b)}, showing how the total energy consumption per hour varies over time. The Figure \ref{fig:label6:(a)} and \ref{fig:label6:(b)} also show that PAR of MADDPG-GCPN is slightly lower than that of DDPG, which is hugely beneficial to operating microgrids reliably. Finally, note that the cost-difference between QP and MADDPG-GCPN is caused by not only the algorithm-difference but also mainly the knowledge-difference. QP is implemented assuming that it knows exactly the future data and system’s dynamics, which is not the case for MADDPG-GCPN.

\section{Conclusion}
In this paper, we proposed a multi-agent actor-critic algorithm with a generative cooperative policy network. The samples generated by the GCPNs of each agent lead other agents to explore over action space and train collaborative policies effectively. We showed the proposed algorithm learned more cooperative policy than existing algorithms. In addition, we implemented the proposed algorithm to the real-world problem of controlling distributed ESSs in microgrids.

Our algorithm has drawbacks in that it requires training additional policies for generating samples. This requires more computation and memory than existing algorithms. In future work, effective approximation methods for computing gradients and effective sampling methods for drawing actions from policies will be studied to address this issue.



\end{document}